\documentclass{desyproc}

\begin{document}
\title{Inclusive Deep-Inelastic Scattering at HERA}

\author{{\slshape Zhiqing Zhang\footnote{On behalf of the H1 and ZEUS Collaborations}}\\[1ex]
Laboratoire de l'Acc\'el\'erateur Lin\'eaire, Univ.\ Paris-Sud 11 and IN2P3/CNRS, Orsay France}

\contribID{259}

\confID{8648}  
\desyproc{DESY-PROC-2014-04}
\acronym{PANIC14} 
\doi  

\maketitle

\begin{abstract}
This contribution covers three recent results on deep-inelastic scattering at HERA: (i) new measurements of the proton longitudinal structure function $F_L$ from H1 and ZEUS experiments,
(ii) a dedicated NC cross section measurement from ZEUS in the region of high Bjorken $x$, and (iii) preliminary combination results of all HERA inclusive data published up to now by H1 and ZEUS, taking into account the experimental correlations between measurements. 
\end{abstract}

\section{Introduction}

At the electron-proton 
$(ep)$ collider HERA, 
the inclusive neutral current (NC) differential cross sections in Bjorken $x$, the virtuality $Q^2$ and inelasticity $y$ are connected with three different structure functions $F_2$, $F_L$ and $xF_3$ as:
\begin{equation}
\tilde{\sigma}_{\rm NC}(x,Q^2,y)\equiv\frac{d^2\sigma_{\rm NC}}{dxdQ^2}\frac{xQ^4}{2\pi \alpha^2}\frac{1}{Y_+}=\left(F_2-\frac{y^2}{Y_+}F_L-\frac{Y_-}{Y_+}xF_3\right)\label{eq:sigma}
\end{equation}
where $Y_\pm =1\pm (1-y)^2$ and the fine structure constant $\alpha=\alpha(Q^2=0)$. The reduced cross section $\tilde{\sigma}$ differs from the full cross section by a kinematic factor. The $F_2$, corresponding to photon exchange, dominates. At high $y$, the $F_L$ term, proportional to the absorption cross section for longitudinally polarized virtual photons by protons, is sizable. At $Q^2\lesssim 1000\,{\rm GeV}^2$, the $xF_3$ term, arising from $Z$ exchange, is small. The similar relation also exists for the charged current (CC) process.

\section{New $F_L$ measurements}



\begin{figure}[htb]
\center
\includegraphics[width=0.495\textwidth]{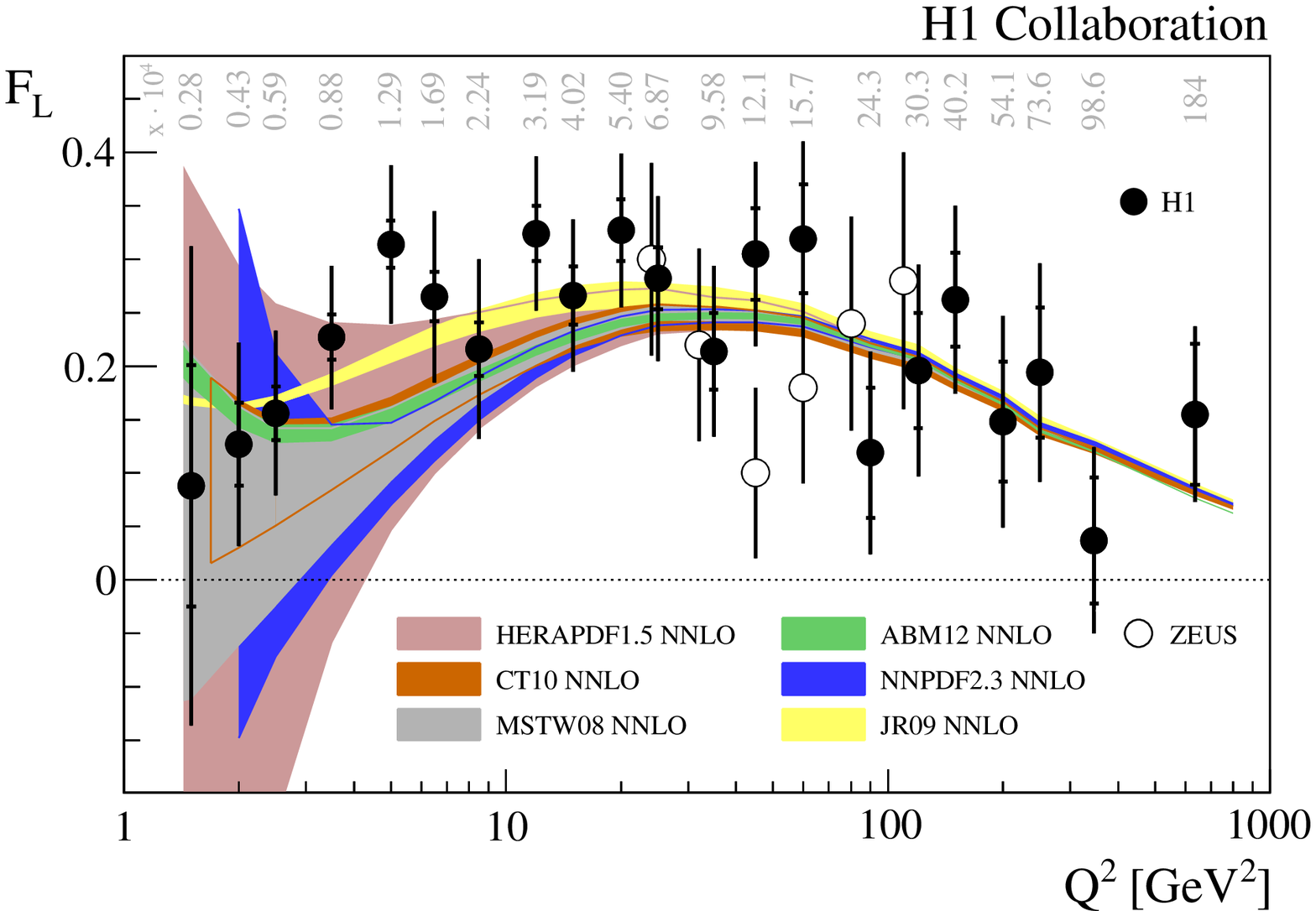}
\includegraphics[width=0.495\textwidth]{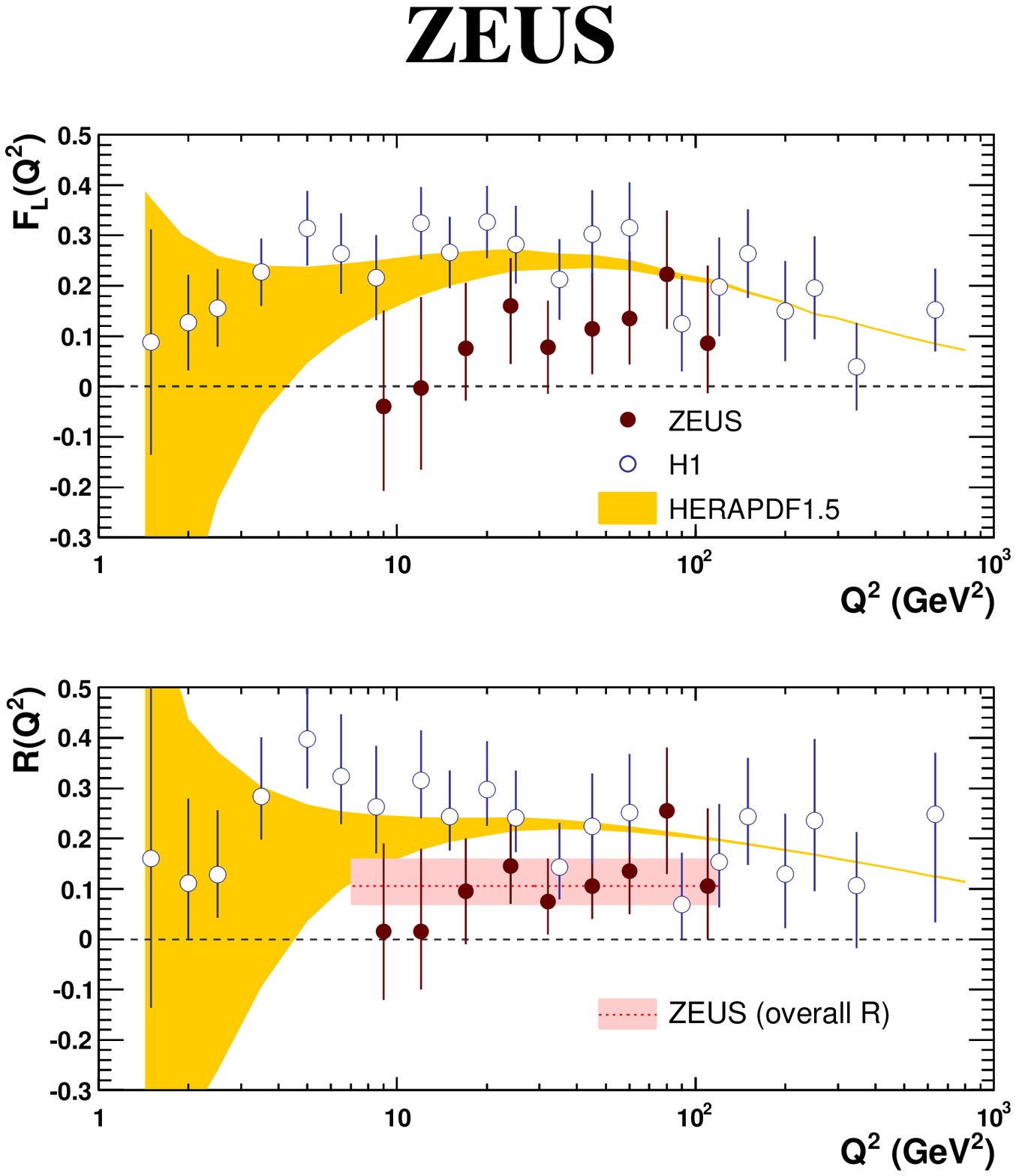}
\caption{Left: new H1 $F_L$ measurement (solid points) in comparison with previous ZEUS measurement (open points) and a few selected NNLO predictions. Right: new ZEUS $F_L$ (a) and $R$ (b) measurements (solid points) in comparison with H1 measurements (open points) and NNLO HERAPDF\,1.5 prediction. The inner error bars represent the statistical uncertainties, the full error bars the total uncertainties. 
The shaded band labelled ``ZEUS (overall R)" represents the 68\% probability interval for the overall $R$.}\label{Fig:fl}
\end{figure}

Using data taken with a lepton beam energy of 27.6\,GeV and two proton beam energies of $E_p=460$ and 575\,GeV corresponding to centre-of-mass energies of 225 and 252\,GeV, respectively, the inclusive NC cross sections have been measured by H1~\cite{h1fl}. The measurements cover the region of $6.5\times 10^{-4}\leq x\leq 0.65$ for $35\leq Q^2\leq 800\,{\rm GeV}^2$ up to the highest accessible inelasticity $y=0.85$. The measurements are used together with previously published H1 data at $E_p=920$\,GeV and lower $Q^2$ data at $E_p=460, 575$ and 920\,GeV to extract 
$F_L$ in the region $1.5\leq Q^2\leq 800\,{\rm GeV}^2$.
The new measurement (Fig.~\ref{Fig:fl}(left)) extends the previous H1 measurements at low and medium $Q^2$ regions~\cite{h1fla,h1flb} to higher $Q^2$ and improves the experimental precision in the region $35\leq Q^2\leq 110\,{\rm GeV}^2$, thus the new measurement supersedes the previous H1 measurements~\cite{h1fla,h1flb}.


Similar measurements have also been performed by ZEUS 
but in a different kinematic region $0.13\leq y\leq 0.75$ and $5\leq Q^2\leq 110\,{\rm GeV}^2$~\cite{zeusflnew}. The new results supersede those in the previous publication~\cite{zeusflold}.
The reduced cross sections were used together with those from the previous ZEUS data collected at $\sqrt{s}=300$\,GeV to extract $F_L$ as well as $F_2$ for 27 values of $x$ and $Q^2$. Relative uncertainties for $F_L$ were in the range of $0.1-0.2$. In addition, $F_L$ 
and the ratio, $R=F_L/(F_2-F_L)$, have also been extracted as a function of $Q^2$ together with an overall value of $R=0.105^{+0.055}_{-0.037}$. The results are shown in Fig.~\ref{Fig:fl}(right). The $F_L$  measurements are lower than but compatible with those in the previous ZEUS and H1 publications and in reasonable agreement with the theoretical prediction.

\section{High $x$ measurement from ZEUS}

\begin{figure}[htb]
\centerline{\includegraphics[width=0.6\textwidth]{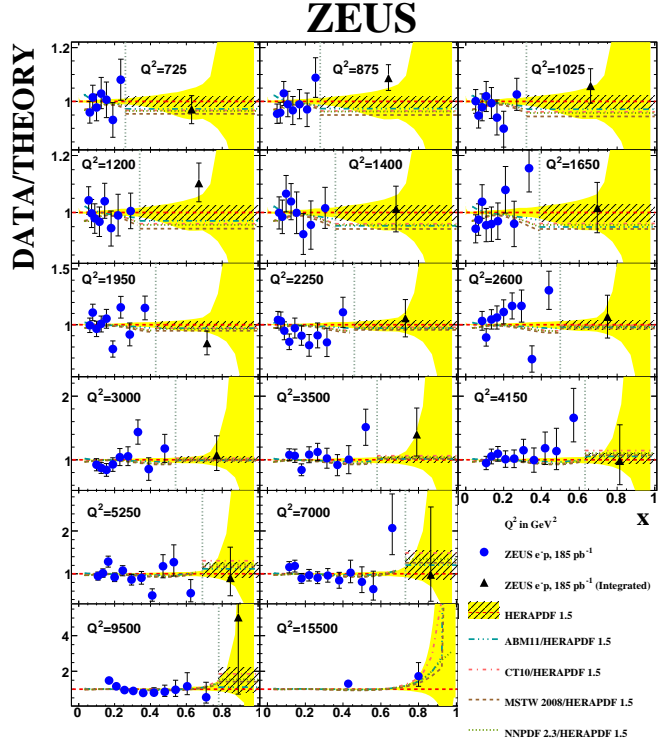}}
\caption{Ratio of the measured NC $e^-p$ cross section over the SM expectation using the HERAPDF1.5 PDFs. The expectation for the integrated high $x$ bin is shown as hatched box.}\label{Fig:hix}
\end{figure}

Motivated by the large uncertainty of parton distribution functions (PDFs) at high $x$, NC $e^\pm p$ cross sections have been measured up to values of $x\simeq 1$ in a dedicated ZEUS analysis using an integrated luminosity of 187\,pb$^{-1}$ of $e^-p$ and 142\,pb$^{-1}$ of $e^+p$ collisions at $\sqrt{s} = 318$\,GeV~\cite{zeushix}. Differential
cross sections in $x$ and $Q^2$ are presented
for $Q^2 \geq 725\,{\rm GeV}^2$ (see Fig.~\ref{Fig:hix} for the ratio of the $e^-p$ measurements over the SM expectations based on a variety of recent PDFs).
An improved reconstruction method and greatly increased
amount of data allow a finer binning in the high-$x$ region of the NC
cross section and lead to a measurement with much improved precision compared
to a similar earlier analysis. The agreement between the measurement and the predictions is
non-trivial as the latter are mostly modeled with a $(1-x)^\beta$ parameterization.

\section{Preliminary combination results of full HERA data}

\begin{figure}[htb]
\center
\includegraphics[width=0.495\textwidth]{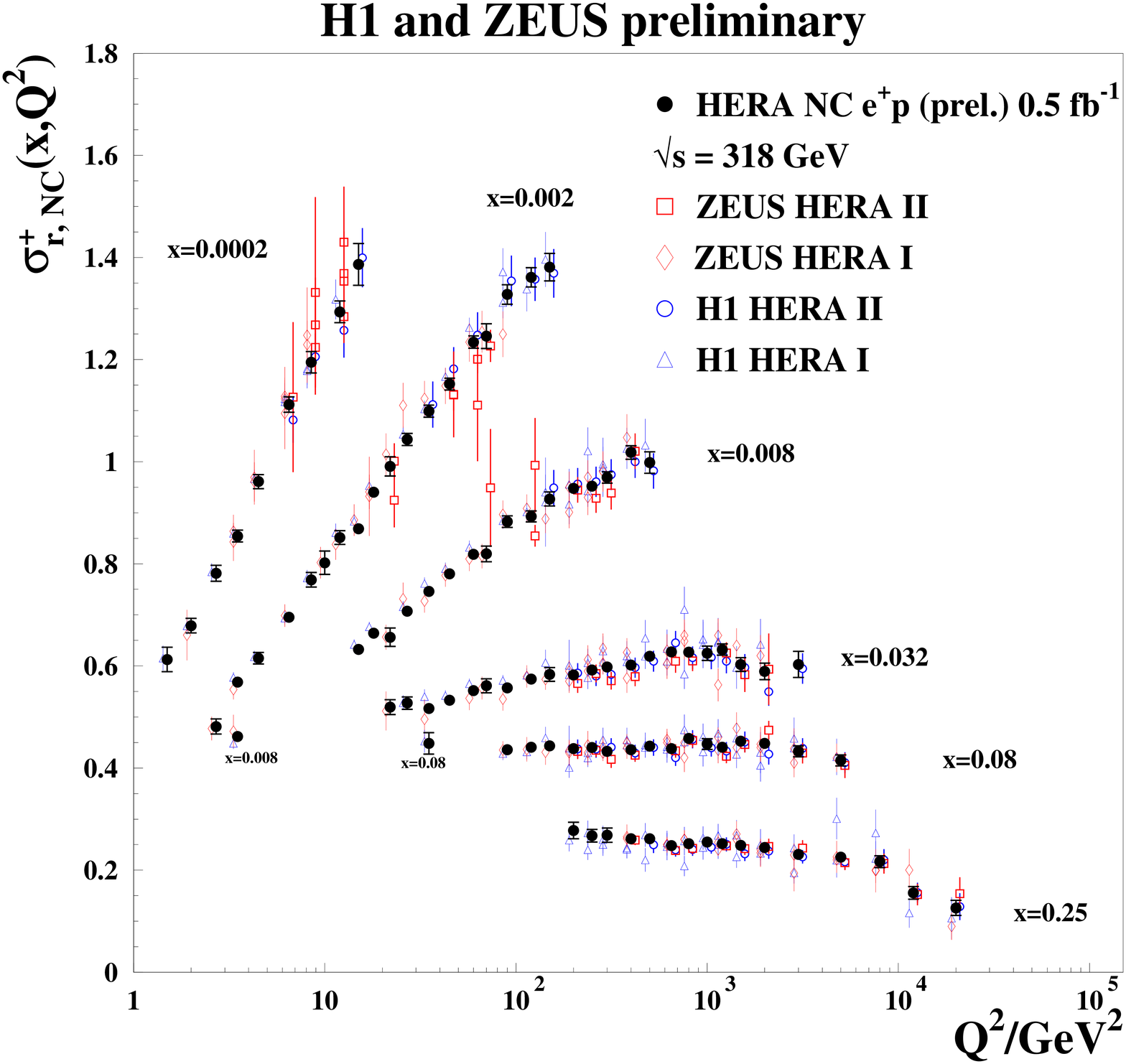}
\includegraphics[width=0.495\textwidth]{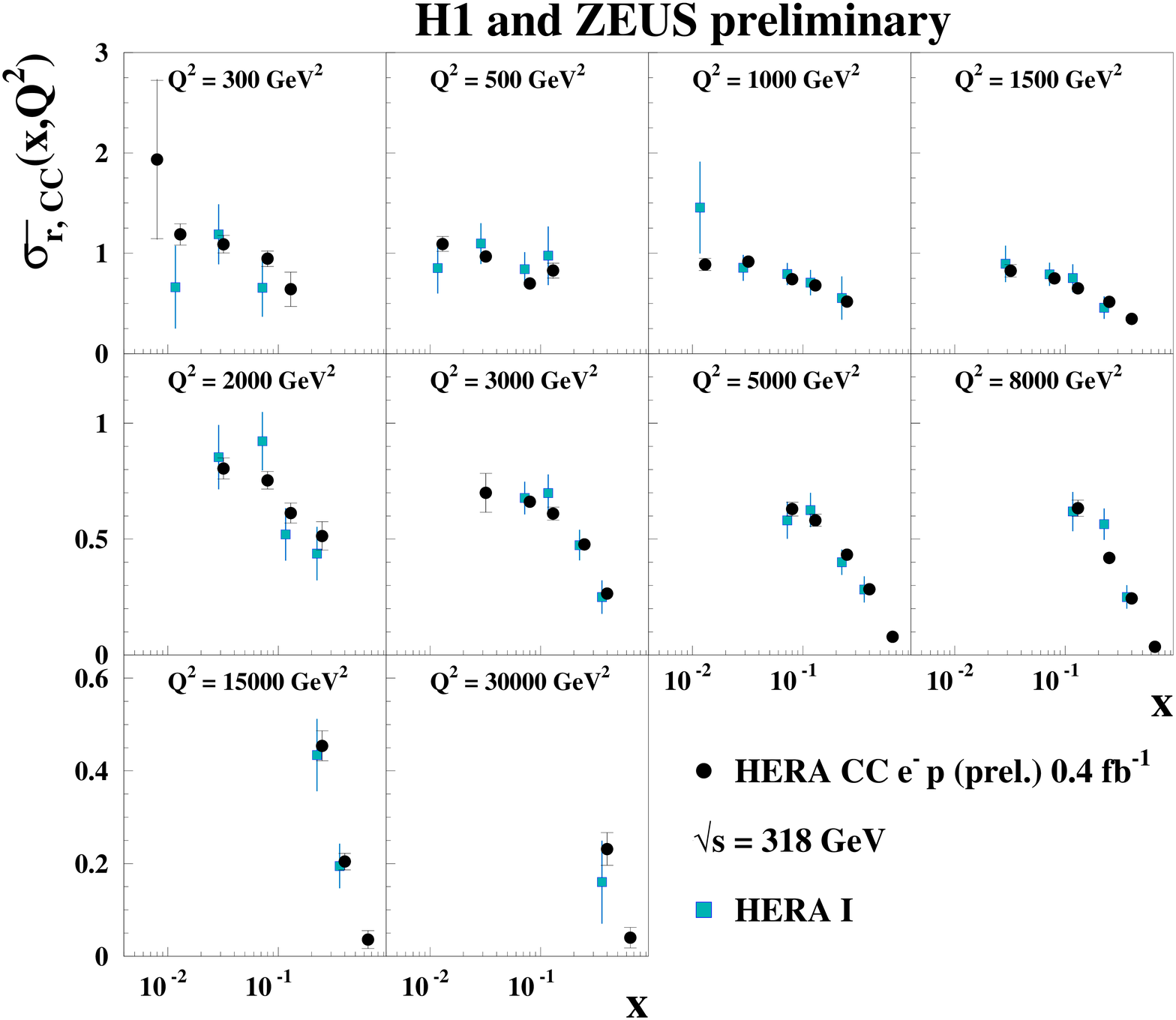}
\caption{Left: Combined NC $e^+p$ reduced cross section as a function of $Q^2$ for six selected $x$-bins compared to the separated H1 and ZEUS data which were input to the combination procedure. Right: Combined CC $e^-p$ reduced cross section as a function of $x$ for 10 $Q^2$ bins in comparison with the results from HERA\,I alone. The error bars represent the total uncertainties.}\label{Fig:comb}
\end{figure}

A preliminary combination is performed of all inclusive deep-inelastic cross sections measured by the
H1 and ZEUS collaborations in NC and CC $e^\pm p$ scattering~\cite{h1zeuscomb}. The data correspond to an integrated luminosity of about 1\,fb$^{-1}$ and span six orders of magnitude
in both $Q^2$ and $x$. They include data taken at proton beam energies of 920, 820, 575 and 460\,GeV. The combination method
used takes the correlations of systematic uncertainties into account, resulting in much improved
accuracy. This is illustrated in Fig.~\ref{Fig:comb} showing part of the combined dataset. The combined data are the inputs for the forthcoming HERAPDF\,2.0 and will also have an important impact on other global PDF sets.

\section{Summary}

Recent results on deep-inelastic scattering at HERA have been presented. The H1 and ZEUS experiments each have determined new measurements of the proton longitudinal structure function $F_L$, making use of the HERA data recorded at reduced centre-of-mass energies. The results are in agreement with each other and with predictions derived from QCD fits. The region of high $x$ is explored in a dedicated measurement by the ZEUS collaboration. All HERA inclusive data published up to now by H1 and ZEUS are combined, taking into account the experimental correlations between measurements. The combined dataset includes measurements of neutral current and charged current cross sections recorded at different centre-of-mass energies, spanning up to six orders of magnitude both in momentum transfer $Q^2$ and in Bjorken $x$. The dataset is superior in precision compared to the previous HERA data combination which included a smaller fraction of the total integrated luminosity collected at HERA. Point-to-point uncorrelated uncertainties better than 1\% are observed in certain kinematic regions.
 

\begin{footnotesize}




\begin{thebibliography}{99}
\bibitem{h1fl}V.~Andreev {\it et al.}  [H1 Collaboration],
  Eur.\ Phys.\ J.\ C {\bf 74} (2014) 2814
  [arXiv:1312.4821 [hep-ex]].
\bibitem{h1fla}F.~D.~Aaron {\it et al.}  [H1 Collaboration],
  Phys.\ Lett.\ B {\bf 665} (2008) 139
  [arXiv:0805.2809 [hep-ex]].
  \bibitem{h1flb}F.~D.~Aaron {\it et al.}  [ H1 Collaboration],
  Eur.\ Phys.\ J.\ C {\bf 71} (2011) 1579
  [arXiv:1012.4355 [hep-ex]].
\bibitem{zeusflnew}H.~Abramowicz {\it et al.}  [ZEUS Collaboration],
  arXiv:1404.6376 [hep-ex].
  \bibitem{zeusflold}S.~Chekanov {\it et al.}  [ZEUS Collaboration],
  Phys.\ Lett.\ B {\bf 682} (2009) 8
  [arXiv:0904.1092 [hep-ex]].
\bibitem{zeushix}H.~Abramowicz {\it et al.}  [ZEUS Collaboration],
  Phys.\ Rev.\ D {\bf 89} (2014) 072007
  [arXiv:1312.4438 [hep-ex]].
\bibitem{h1zeuscomb}H1 and ZEUS Collaborations, {\it Combined measurement of inclusive $e^\pm p$ scattering cross sections at HERA}, H1prelim-14-041, ZEUS-prel-14-005.

\end{thebibliography}
%

\end{footnotesize}


\end{document}